\documentclass[12pt]{article}
\begin{document}
\newcommand{\beq}{\begin{equation}}
\newcommand{\eeq}{\end{equation}}
\newcommand{\beqa}{\begin{eqnarray}}
\newcommand{\eeqa}{\end{eqnarray}}
\newcommand{\beqar}{\begin{eqnarray*}}
\newcommand{\eeqar}{\end{eqnarray*}}
\newcommand{\al}{\alpha}
\newcommand{\be}{\beta}
\newcommand{\del}{\delta}
\newcommand{\D}{\Delta}
\newcommand{\eps}{\epsilon}
\newcommand{\ga}{\gamma}
\newcommand{\Ga}{\Gamma}
\newcommand{\ka}{\kappa}
\newcommand{\inn}{\!\cdot\!}
\newcommand{\h}{\eta}
\newcommand{\kk}{\varphi}
\newcommand\F{{}_3F_2}
\newcommand{\la}{\lambda}
\newcommand{\La}{\Lambda}
\newcommand{\na}{\nabla}
\newcommand{\Om}{\Omega}
\newcommand{\p}{\phi}
\newcommand{\sig}{\sigma}
\renewcommand{\t}{\theta}
\newcommand{\z}{\zeta}
\newcommand{\ssc}{\scriptscriptstyle}
\newcommand{\eg}{{\it e.g.,}\ }
\newcommand{\ie}{{\it i.e.,}\ }
\newcommand{\labell}[1]{\label{#1}} 
\newcommand{\reef}[1]{(\ref{#1})}
\newcommand\prt{\partial}
\newcommand\veps{\varepsilon}
\newcommand\ls{\ell_s}
\newcommand\cF{{\cal F}}
\newcommand\cA{{\cal A}}
\newcommand\cS{{\cal S}}
\newcommand\cH{{\cal H}}
\newcommand\cC{{\cal C}}
\newcommand\cL{{\cal L}}
\newcommand\cG{{\cal G}}
\newcommand\cI{{\cal I}}
\newcommand\cl{{\iota}}
\newcommand\cP{{\cal P}}
\newcommand\cV{{\cal V}}
\newcommand\cg{{\it g}}
\newcommand\cR{{\cal R}}
\newcommand\cB{{\cal B}}
\newcommand\cO{{\cal O}}
\newcommand\tcO{{\tilde {{\cal O}}}}
\newcommand\bz{\bar{z}}
\newcommand\bw{\bar{w}}
\newcommand\hF{\hat{F}}
\newcommand\hA{\hat{A}}
\newcommand\hT{\hat{T}}
\newcommand\htau{\hat{\tau}}
\newcommand\hD{\hat{D}}
\newcommand\hf{\hat{f}}
\newcommand\hg{\hat{g}}
\newcommand\hp{\hat{\phi}}
\newcommand\hh{\hat{h}}
\newcommand\ha{\hat{a}}
\newcommand\hQ{\hat{Q}}
\newcommand\hP{\hat{\Phi}}
\newcommand\hb{\hat{b}}
\newcommand\hc{\hat{c}}
\newcommand\hd{\hat{d}}
\newcommand\hS{\hat{S}}
\newcommand\hX{\hat{X}}
\newcommand\tL{\tilde{\cal L}}
\newcommand\hL{\hat{\cal L}}
\newcommand\tG{{\widetilde G}}
\newcommand\tg{{\widetilde g}}
\newcommand\tphi{{\widetilde \phi}}
\newcommand\tPhi{{\widetilde \Phi}}
\newcommand\te{{\tilde e}}
\newcommand\tk{{\tilde k}}
\newcommand\tf{{\tilde f}}
\newcommand\tF{{\widetilde F}}
\newcommand\tK{{\widetilde K}}
\newcommand\tE{{\widetilde E}}
\newcommand\tpsi{{\tilde \psi}}
\newcommand\tX{{\widetilde X}}
\newcommand\tD{{\widetilde D}}
\newcommand\tO{{\widetilde O}}
\newcommand\tS{{\tilde S}}
\newcommand\tB{{\widetilde B}}
\newcommand\tA{{\widetilde A}}
\newcommand\tT{{\widetilde T}}
\newcommand\tC{{\widetilde C}}
\newcommand\tV{{\widetilde V}}
\newcommand\thF{{\widetilde {\hat {F}}}}
\newcommand\Tr{{\rm Tr}}
\newcommand\tr{{\rm tr}}
\newcommand\STr{{\rm STr}}
\newcommand\M[2]{M^{#1}{}_{#2}}
\parskip 0.3cm

\vspace*{1cm}

\begin{center}
{\bf \Large   Ricci curvature corrections to type II supergravity     }

\vspace*{1cm}

{  Mohammad R. Garousi\footnote{garousi@ferdowsi.um.ac.ir} }\\
\vspace*{1cm}
{ Department of Physics, Ferdowsi University of Mashhad,\\ P.O. Box 1436, Mashhad, Iran}
\\
\vspace{2cm}

\end{center}

\begin{abstract}
\baselineskip=18pt

We examine the  NS-NS corrections to the type II supergravity  given by Gross and Sloan,  under  the
 linear  T-duality transformations.  We show  that the gravity and the B-field couplings are invariant under  off-shell T-duality, whereas the dilaton couplings are invariant under on-shell T-duality. Requiring  the compatibility of the dilaton couplings with off-shell T-duality   as a guiding principle, we extend them to include the appropriate Ricci curvature and the divergence of the B-field strength.    
 

\end{abstract}
Keywords: T-duality, effective action

\setcounter{page}{0}
\setcounter{footnote}{0}
\newpage
\section{Introduction  } \label{intro}

T-duality is one of the most fascinating dualities of superstring theory \cite{Kikkawa:1984cp,TB,Giveon:1994fu,Alvarez:1994dn} . It relates  type IIA superstring theory compactified on a circle with radius $\rho$ to  type IIB superstring theory compactified on another circle with radius $\alpha'/\rho$. At low energy, this duality relates  type IIA supergravity  to  type IIB  supergravity. The   NS-NS sector of these theories, \ie
\beqa
S&=&\frac{1}{2\kappa^2}\int d^{10}x e^{-2\Phi}\sqrt{-G}\bigg[R+4(\prt\Phi)^2-\frac{1}{3}H^2\bigg]\labell{super}
\eeqa
is invariant under the T-duality. The higher derivative corrections to this action in which we are interested, may be found by the combination of the S-matrix calculation \cite{Green:1981xx,Green:1981ya} and the T-duality \cite{Garousi:2009dj,Garousi:2010ki,Becker:2010ij,Garousi:2010rn,Garousi:2010bm,Garousi:2011ut,Becker:2011bw,Becker:2011ar,Velni:2012sv,McOrist:2012yc}. 

The higher derivative corrections to this action start at order $\alpha'^3$ \cite{Gross:1986iv}. For  gravity, they involve couplings with structure $R^4$ where $R$ refers to the Riemann, the Ricci or the scalar curvatures. For B-field, they involve couplings with structure $(\prt H)^4$, $(\prt H)^2H^4$ or $H^8$. Similarly for dilaton and for mixed couplings. These higher derivative couplings appear in the contact terms of the corresponding string theory S-matrix elements at order $\alpha'^3$. For example, the couplings with structure $H^8$ appear in the contact terms of the S-matrix element of eight Kalb-Ramond vertex operators. The evaluation of this eight-point function and the extraction of  its contact terms, however,  are  extremely difficult tasks. So one has to take advantage of  the symmetries/dualities  of the effective field theory to find such higher derivative couplings. Supersymmetry is able to determine all  $\alpha'^3$-couplings   up to field redefinitions \cite{Paban:1998ea,Green:1998by}. It can also fix the moduli-dependence of the type IIB theory \cite{Green:1998by}.

Even for the gravity couplings which can be extracted from the four-point function, there is no unique expression for the $R^4$ couplings. The $R^4$ couplings which reproduce the sigma model beta function \cite{Grisaru:1986vi,Freeman:1986zh} and the particular $R^4$ couplings extracted from the S-matrix element \cite{Myers:1987qx,Gross:1986mw}  are identical up to some terms containing the Ricci and scalar curvatures  \cite{Myers:1987qx}. These terms can be absorbed by the supergravity \reef{super} in the field redefinition of the metric $G\rightarrow G+\delta G$ \cite{Gross:1986iv}.  The field redefinitions, however, change the standard forms of the duality transformations to non-standard forms which receive higher derivative corrections. The field redefinition ambiguity   may then  be fixed by requiring the effective action to be invariant under the standard form of the duality transformations. 
We are interested in the   effective action which is compatible with the standard form of the T-duality transformations \cite{TB}.

The T-duality at the linear order appears in the S-matrix elements through the appropriate Ward identity \cite{Garousi:2011we,Velni:2012sv}. The  duality  in this case  is on-shell because  one has to use various on-shell relations to verify the Ward identity \cite{Garousi:2011we,Velni:2012sv}. The  T-duality of the effective action, however,  must be off-shell because the effective action is an off-shell object. To study the T-duality   at the most simple level, one has to examine the compatibility of the effective action with the  linear T-duality. At the more advance level, one has to extend the linear T-duality to full nonlinear T-duality.  The appropriate framework  for this part is   the double field theory formalism in which the fields depend both on the usual spacetime coordinates and on the winding coordinates \cite{Hull:2009mi,Hohm:2010jy}. In this formalism the metric and the B-field can appear in a generalized metric which includes the linear and the quadratic order of B-field \cite{Hohm:2010pp}. The effective action  in terms of the generalized metric then would include all couplings at order $\alpha'^3$ in a unique form. Therefore, it seems there are two steps to find the higher derivative effective action. The first step in which we are interested in this paper, is to find the appropriate couplings of four NS-NS fields which are invariant under off-shell linear T-duality. The second steps would be  to rewrite the couplings in terms of the generalized metric in which the quadratic order of B-field is set to zero. Releasing this latter condition might incorporate  automatically the higher order fields.  

   The Riemann curvature couplings  at order $\alpha'^3$ have been found in \cite{ Gross:1986iv} by analyzing the sphere-level four-graviton scattering amplitude in type II superstring theory. The result in the eight-dimensional transverse space of the light-cone formalism, is a polynomial in the Riemann curvature tensors 
\beqa
\cL&\sim& t^{i_1\cdots i_8}t_{j_1\cdots j_8}R_{i_1i_2}{}^{j_1j_2}\cdots R_{i_7i_8}{}^{j_7j_8}+\cdots \labell{Y1}
\eeqa
where $t^{\{\cdots\}}$ is a tensor in eight dimensions which includes the eight-dimensional Levi-Civita tensor \cite{Gross:1986iv} and dots represent terms containing the Ricci and scalar curvature tensors. They can not be captured by the four-graviton scattering amplitude as they are zero on-shell. 
The above $SO(8)$ invariant  Lagrangian has been extended to the following Lorentz invariant form   in \cite{Grisaru:1986vi,Freeman:1986zh}:
\beqa
\cL&\sim&   R_{hmnk}R_p{}^{mn}{}_qR^{hrsp}R^q{}_{rs}{}^k+\frac{1}{2}R_{hkmn}R_{pq}{}^{mn}R^{hrsp}R^q{}_{rs}{}^k+\cdots\labell{Y2}
\eeqa
where   dots represent the specific form of the  off-shell  Ricci and scalar curvature couplings which reproduce the sigma model beta function \cite{Grisaru:1986vi,Freeman:1986zh}.  The B-field and the dilaton have been added into this action by requiring the consistency of the action with the   linear T-duality \cite{Garousi:2012yr}.  The resulting theory  however satisfies the duality only after using on-shell relations. In particular, under the dimensional reduction on a circle the couplings \reef{Y2} produce the following coupling: 
\beqa
4  R_{ k yn y } R^{ n}{}_{ yq y } R_{ r ys y }R^{ q rk s }\eta^{yy}\eta^{yy}\eta^{yy}\labell{non}
\eeqa
where $y$ is the  Killing index. This couping which is not invariant under the linear T-duality, is zero after using on-shell relations \cite{Garousi:2012yr}.

An alternative expression for the  $SO(8)$ invariant Lagrangian \reef{Y1} has been found by Gross and Sloan  in \cite{Gross:1986mw} by expanding the kinematic factors $t^{\{\cdots\}}$. Moreover,  the $B$-field and the dilaton have been included in this Lagrangian by extending the Riemann curvature to an expression which includes the first derivative of $H$ and the second derivative of the dilaton \cite{Gross:1986mw}.   In this paper we would like to find the specific form of the off-shell couplings containing the   Ricci curvature  and the divergence of $H$   which make this effective action to be invariant under off-shell linear T-duality.

The outline of the paper is as follows: We begin in section 2 by reviewing the linear T-duality transformation of the linearized Riemann and Ricci curvatures. In section 3, we observe  that while the gravity and the B-field couplings in the eight-dimensional Lagrangian  found in \cite{Gross:1986mw} can naturally be extended to ten-dimensional spacetime, the dilaton couplings are valid only in eight dimensions. In this section we extend the eight-dimensional Lagrangian to a Lagrangian which is valid in ten dimensions.  In section 4, we  observe that the prescription given by Gross and Sloan  for finding the dilaton coupling is the same as transforming  the ten-dimensional Einstein frame to the string frame. Using this  observation,  we transform the Einstein frame  ten-dimensional   Lagrangian    to the   string frame and  show that the   action is invariant under the on-shell linear T-duality. In section 5, we  extend it to off-shell T-duality by including the specific off-shell couplings containing the Ricci curvature and the divergence of the B-field strength. In section 6, we briefly discuss our results.

\section{T-duality}

The full set of nonlinear T-duality transformations   have been found in \cite{TB}.   When  the T-duality transformation acts along the Killing coordinate $y$,  the massless NS-NS fields   transform as:
\beqa
e^{2\tPhi}=\frac{e^{2\Phi}}{G_{yy}}&;& 
\tG_{yy}=\frac{1}{G_{yy}}\nonumber\\
\tG_{\mu y}=\frac{B_{\mu y}}{G_{yy}}&;&
\tG_{\mu\nu}=G_{\mu\nu}-\frac{G_{\mu y}G_{\nu y}-B_{\mu y}B_{\nu y}}{G_{yy}}\nonumber\\
\tB_{\mu y}=\frac{G_{\mu y}}{G_{yy}}&;&
\tB_{\mu\nu}=B_{\mu\nu}-\frac{B_{\mu y}G_{\nu y}-G_{\mu y}B_{\nu y}}{G_{yy}}\labell{nonlinear}
\eeqa
where $\mu,\nu$ denote any coordinate directions other than $y$. In above transformation the metric is given in the string frame. If $y$ is identified on a circle of radius $\rho$, \ie $y\sim y+2\pi \rho$, then after T-duality the radius becomes $\tilde{\rho}=\alpha'/\rho$. The string coupling $g=e^{\phi_0}$ is also shifted as $\tilde{g}=g\sqrt{\alpha'}/\rho$.

We would like to study the T-duality of the $\alpha'^3$-order couplings of type II superstring theory which involve four NS-NS fields, so we need the above transformations at the linear order. Assuming that the NS-NS  fields are small perturbations around the background, \ie
\beqa
G_{\mu\nu}&=&\eta_{\mu\nu}+2\kappa h_{\mu\nu}\,;\,\,G_{yy}\,=\,\frac{\rho^2}{\alpha'}(1+2\kappa h_{yy})\,;\,\,G_{\mu y}\,=\,2\kappa h_{\mu y}\nonumber\\
B_{\mu\nu}&=&2\kappa b_{\mu\nu}\,;\,B_{\mu y}\,=\,2\kappa b_{\mu y}\,;\,\,
\Phi\,=\,\phi_0+\sqrt{2}\kappa \phi\labell{perturb}
\eeqa
  the   transformations \reef{nonlinear} take the following linear form for the perturbations:
\beqa
&&
 \sqrt{2}\tilde{\phi}=\sqrt{2}\phi - h_{yy},\,\tilde{h}_{yy}=-h_{yy},\, \tilde{h}_{\mu y}=b_{\mu y},\, \tilde{b}_{\mu y}=h_{\mu y},\,\tilde{h}_{\mu\nu}=h_{\mu\nu},\,\tilde{b}_{\mu\nu}=b_{\mu\nu}\labell{linear}
\eeqa
  
The effective action at order $\alpha'^3$ includes the Riemann curvature and the derivative of  the B-field strength $H=\frac{1}{2}dB$. So we need the transformation of these fields under the linear T-duality.  The transformation of the linearized Riemann curvature tensor  when it carries zero, one and two Killing indices are, respectively, \cite{Garousi:2012yr}  
\beqa
R_{abcd}&\leftrightarrow & R_{abcd}\nonumber\\
R_{abc y}&\leftrightarrow & -  H_{ab y,c}\labell{TR}\\
R_{ayby}&\leftrightarrow &-R_{ayby}\nonumber
\eeqa
where the linearized Riemann curvature and $H$ are
\beqa
R_{abcd}&=& \kappa(h_{ad,bc}+h_{bc,ad}-h_{ac,bd}-h_{bd,ac})\nonumber\\
H_{abc}&=&\kappa(b_{ab,c}+b_{ca,b}+b_{bc,a})\labell{RH}
\eeqa
where as usual  the comma represents partial differentiation. The field strength $H$ is also  invariant when it carries no Killing index.

As we will see in the next section the effective action in the string frame involves the second derivative of the dilaton which is not invariant under the T-duality. It must be combined with the Ricci curvature to become invariant \cite{Garousi:2012yr}. The transformation of the Ricci curvature when it carries zero, one and two $y$ indices are, respectively, \cite{Garousi:2012yr} 
\beqa
R_{ab}+2\sqrt{2}\kappa \phi_{,ab}&\leftrightarrow &R_{ab}+2\sqrt{2}\kappa \phi_{,ab}\labell{TR2}\\
 R_{ay}&\leftrightarrow &-H_{ay} \nonumber\\
R_{yy}&\leftrightarrow& -R_{yy}\nonumber 
\eeqa
where $H_{ab}$ is the divergence of the B-field strength, \ie $H_{ab} \equiv H_{abc}{}^{;c}$ where the semicolon represents covariant differentiation.

In a T-duality invariant theory, all the couplings in the dimensional reduction must be invariant under the above T-duality transformations. So to extend a coupling to a set of couplings which   
  are invariant under linear T-duality,    we first use the dimensional reduction to reduce the 10-dimensional couplings to 9-dimensional couplings, \ie separate the  indices  along and orthogonal to $y$,     and then apply the  above T-duality transformations.  If the original coupling is not invariant under the T-duality, one must add new terms to make them invariant.  
  
Before ending this section, let us compare the above  method for finding the T-duality invariant couplings in the  spacetime, and the method used in \cite{Garousi:2009dj} for finding the T-duality invariant couplings on the D-brane world volume. In both cases, one first uses  the dimensional reduction to separate the indices along and orthogonal to the Killing $y$-direction, then uses the linear T-duality transformations. After the T-duality has been performed, the indices must be complete.  In a T-duality invariant theory in spacetime, the completion of the $y$-index trivially gives the original theory because the $y$-index is a spacetime index before and after T-duality. However, in a D-brane theory the world volume Killing index $y$ becomes the transverse index after T-duality. Hence, the completion of the transverse indicies gives a nontrivial constraint on the D-brane couplings \cite{Garousi:2009dj}.   

\section{10-dimensional action  }

The   Lagrangian \reef{Y1} has the eight-dimensional Levi-Civita tensor in its kinematic factor $t^{i_1\cdots i_8}$ \cite{Gross:1986iv} so this form of Lagrangian is valid only in eight dimensions. The   dilaton  and the B-field fluctuations have been added into this Lagrangian by the following replacement \cite{Gross:1986mw}: 
\beqa
R_{ab}{}^{cd}\rightarrow \bar{R}_{ab}{}^{cd}&\equiv &R_{ab}{}^{cd}+  e^{-\phi_0/2}H_{ab}{}^{[c;d]}-\frac{\kappa}{\sqrt{8}}\eta_{[a}{}^{[c}\phi_{;b]}{}^{d]}\labell{trans}
\eeqa
where the bracket notation is $H_{ab}{}^{[c;d]}=H_{ab}{}^{c;d}-H_{ab}{}^{d;c}$ and $\phi_0$ is the constant dilaton background. The Lagrangian \reef{Y1} along with the  replacement \reef{trans} reproduces exactly the string theory S-matrix elements.  

In order to extend the Lagrangian \reef{Y1} to ten dimensions, the first step is to expand  the kinematic factor $t^{i_1\cdots i_8}$ which  has been done in  \cite{Gross:1986mw}  
\beqa
\frac{\gamma e^{-3\phi_0/2}}{\kappa^2}\cL_1(\bar{R})&\!\!\!\!=\!\!\!\!&\frac{\gamma e^{-3\phi_0/2}}{\kappa^2}\big[\bar{R}_{hkmn}\bar{R}^{krnp}\bar{R}_{rs}{}^{qm}\bar{R}^{sh}{}_{pq}+\frac{1}{2}\bar{R}_{hkmn}\bar{R}^{krnp}\bar{R}_{rspq}\bar{R}^{shqm}\labell{Y3}\\
&&\qquad\qquad -\frac{1}{2}\bar{R}_{hkmn}\bar{R}^{krmn}\bar{R}_{rspq}\bar{R}^{shpq} -\frac{1}{4}\bar{R}_{hkmn}\bar{R}^{krpq}\bar{R}_{rs}{}^{mn}\bar{R}^{sh}{}_{pq}\nonumber\\
&&\qquad\qquad+\frac{1}{16}\bar{R}_{hkmn}\bar{R}^{khpq}\bar{R}^{rsmn}\bar{R}_{srpq}+\frac{1}{32}\bar{R}_{hkmn}\bar{R}^{khmn}\bar{R}_{rspq}\bar{R}^{srpq}\bigg]+\cdots\nonumber
\eeqa
where $\gamma=\frac{\alpha'^3}{2^5} \z(3)$. We have called this Lagrangian $\cL_1$ because as we will see shortly there are two other Lagrangians   in   ten dimensions.

The gravity and the B-field parts of the above  Lagrangian are valid   in ten dimensions, however, the   dilaton part is  valid only in eight  dimensions. To clarify this point we first   write the replacement \reef{trans} for the linear perturbations. Using the equation \reef{RH}, one can write \reef{trans} as 
\beqa
h_{ab}\rightarrow h_{ab}+e^{-\phi_0/2}b_{ab}+\eta_{ab}\phi/\sqrt{8}\labell{ltrans}
\eeqa
  This gives the following transformations for the expression $h_{ab}h ^{ba}$ in eight   dimensions 
  \beqa
  h_{ab}h ^{ba}\rightarrow  h_{ab}h ^{ba}+e^{-\phi_0}b_{ab}b^{ba}+\phi^2\labell{h2} 
  \eeqa
 This is what one finds also in the string theory scattering amplitude \cite{Garousi:2012yr}. That is, if we write the graviton polarization tensor by $\veps^{ab}$, then the term corresponding to $h_{ab}h ^{ba}$ is $\Tr[\veps_1\inn\veps_2]$. The prescription given in the string theory for calculating  the dilaton amplitude from the graviton amplitude,  gives the following replacement \cite{Garousi:2012yr}:
 \beqa
 \Tr[\veps_1\inn\veps_2]\rightarrow \phi_1\phi_2
\eeqa
in   the scattering amplitude,  which is consistent with the   replacement \reef{h2} in the field theory. However, the replacement \reef{ltrans} gives the following result in ten dimensions:
\beqa
  h_{ab}h ^{ba}\rightarrow  h_{ab}h ^{ba}+e^{-\phi_0}b_{ab}b^{ba}+\frac{5}{4}\phi^2
\eeqa
which is not consistent with the string theory scattering amplitude. Note that the difference between eight- and ten-dimensional couplings happen  only for the couplings in \reef{Y3} which contain the expression $h_{ab}h ^{ba}$. All other  couplings   are the same in both eight and ten dimensions.

To extend the Lagrangian \reef{Y3}   to ten dimensions, we rewrite the replacement \reef{h2} as
\beqa
  h_{ab}h ^{ba}\rightarrow  h_{ab}h ^{ba}+e^{-\phi_0}b_{ab}b^{ba}+\frac{5}{4}\phi^2-\frac{1}{4}\phi^2\labell{h210} 
  \eeqa
The couplings in \reef{Y3} corresponding to the first three terms above are reproduced by the ten-dimensional extension of the replacement \reef{trans}, and the couplings corresponding to the last term above must be subtracted from the ten-dimensional theory. To find these latter couplings, we have to find the couplings in \reef{Y3} which contain $h_{ab}h ^{ba}$ and then replace $h_{ab}h ^{ba}$ in them by $\frac{1}{4}\phi^2$. 

A simple way to find the couplings in \reef{Y3} which contain $h_{ab}h ^{ba}$ is to use the dimensional reduction on a circle and look for the terms containing $h_{yy}h^{yy}$. On the other hand, the dimensional reduction of the linearized Riemann curvature \reef{RH} with two killing indices is $R_{ayby}=-\frac{\kappa \rho^2}{\alpha'}h_{yy,ab}$. Therefore, we have to find the couplings in the dimensional reduction of   \reef{Y3} which contain two such Riemann curvatures. They are 
\beqa
 &&\frac{\gamma e^{-3\phi_0/2}}{\kappa^2} \bigg[2 \bar{R}_{ m q  r  s } \bar{R}^{ p q  h s } R_{ hy}{}^{my} R^{ ry}{}_{py}+2 \bar{R}_{m  q  h  s } \bar{R}^{ p q r s } R^{ hymy} R_{ rypy}\nonumber\\
&&\qquad\qquad+2 \bar{R}_{  h  q m {p} } \bar{R}^{  q r m {p}} R^{ hy sy} R_{ ry sy}+\frac{1}{4} \bar{R}_{ h p m {r} } \bar{R}^{ h p m {r} } R_{ qy}{}^{sy} R^{ qy}{}_{sy}\bigg]\labell{RP1}
\eeqa
The terms that must be subtracted are then given by the above couplings in which $h_{yy}h^{yy}$ is replaced by $\frac{1}{4}\phi^2$. For example, in the first term one must replace $R_{ hy}{}^{my} R^{ ry}{}_{py}$ with $\frac{\kappa^2}{4}\phi_{ ;h}{}^{m} \phi^{ ;r}{}_{p}$.  Therefore, the following   Lagrangian must be added to the ten-dimensional extension of \reef{Y3}:
\beqa
\frac{\gamma e^{-3\phi_0/2}}{\kappa^2}\cL_2(\bar{R},\phi)&=&\frac{\gamma e^{-3\phi_0/2}}{\kappa^2}(-\frac{\kappa^2}{4})\bigg[ 2 \bar{R}_{ m q  r  s } \bar{R}^{ pq  h s } \phi_{ ;h}{}^{m} \phi^{ ;r}{}_{p}+2 \bar{R}_{m  q  h  s } \bar{R}^{ p q r s } \phi^{ ;hm} \phi_{ ;rp}\nonumber\\
&& +2 \bar{R}_{  h {q}m  p } \bar{R}^{  q r m {p}} \phi^{ ;h}{}_{s} \phi_{ ;r}{}^s+\frac{1}{4} \bar{R}_{ h p m {r} } \bar{R}^{ h p m {r} } \phi_{ ;q}{}^{s} \phi^{ ;q}{}_s\bigg]\labell{L2}
\eeqa
We have explicitly checked that the sum of the two-dilaton couplings in the above equation and   in the ten-dimensional extension of  \reef{Y3},    produces exactly the two-dilaton couplings in the eight-dimensional Lagrangian \reef{Y3}.

The eight-dimensional Lagrangian \reef{Y3} has also terms containing  $(h_{ab}h ^{ba})^2$. The replacement \reef{trans}   in eight dimensions gives
 \beqa
  (h_{ab}h ^{ba})^2\rightarrow  (h_{ab}h ^{ba}+e^{-\phi_0}b_{ab}b^{ba}+\phi^2)^2\labell{h22} 
  \eeqa
which is consistent with the string scattering amplitude \cite{Garousi:2012yr}. This term can also be rewritten as
\beqa
  (h_{ab}h ^{ba})^2\rightarrow  (h_{ab}h ^{ba}+e^{-\phi_0}b_{ab}b^{ba}+\frac{5}{4}\phi^2)^2-\frac{1}{2}\phi^2(h_{ab}h ^{ba}+e^{-\phi_0}b_{ab}b^{ba}+\frac{5}{4}\phi^2)+\frac{1}{16}\phi^4\labell{h2210} 
  \eeqa
The first part again corresponds to the couplings in the ten-dimensional extension of \reef{Y3}, the second part corresponds to the couplings given by \reef{RP1} and the last term   corresponds to yet another Lagrangian which must be added to   the ten-dimensional Lagrangians  \reef{Y3} and \reef{L2}. 

To catch this  Lagrangian, one may first find the couplings in \reef{Y3} which contain $(h_{ab}h ^{ba})^2$ and then replace $(h_{ab}h ^{ba})^2$ in them by $\frac{1}{16}\phi^4$. We   use this replacement in the couplings containing $(h_{yy}h^{yy})^2$ in the dimensional reduction of \reef{Y3}. Note that the couplings in \reef{Y3} with structure $h_{ab}h^{bc}h_{cd}h^{da}$ do not survive in the dimensional reduction \cite{Garousi:2012yr}.    The couplings in the dimensional reduction of   \reef{Y3} which contain four  Riemann curvatures   with two Killing indices are the following:
 \beqa
\frac{\gamma e^{-3\phi_0/2}}{\kappa^2} \bigg[R_{ ky}{}^{ny} R^{ ky}{}_{py} R_{ sy}{}^{ny} R^{ sy}{}_{py}+\frac{ 1}{2} (R_{ ky}{}^{ny}R_{ny}{}^{ky} )^2\bigg]\labell{p41}
\eeqa
 The terms that must be subtracted are then given by the above couplings in which $(h_{yy}h^{yy})^2$ is replaced by $\frac{1}{16}\phi^4$.

The direct way to observe the above four-dilaton couplings is to find the couplings in \reef{Y3} which has $(\eta_{ab}\eta^{ab})^2$. They are given by the following expression:  
\beqa
\kappa^2\gamma e^{-3\phi_0/2}\bigg[\frac{3}{2}(\phi_{;hm}\phi^{;hm})^2-\phi_{;hn}\phi^{;hq}\phi_{;qr}\phi^{;rn}\bigg]\frac{(\eta_{ab}\eta^{ab})^2}{64}
\eeqa
The factor  $\frac{(\eta_{ab}\eta^{ab})^2}{64}$ is 1 in eight dimensions and is   $(5/4)^2$ in ten dimensions. Moreover, the ten-dimensional  Lagrangian \reef{L2} produces similar  four-dilaton couplings   
\beqa
-\kappa^2\gamma e^{-3\phi_0/2}\bigg[\frac{3}{2}(\phi_{;hm}\phi^{;hm})^2-\phi_{;hn}\phi^{;hq}\phi_{;qr}\phi^{;rn}\bigg]\frac{\eta_{ab}\eta^{ab}}{16}
\eeqa
The factor  $\frac{\eta_{ab}\eta^{ab}}{16}$ is $5/8$ in ten dimensions. Writing  the factor 1 in the eight dimensions as $(5/4)^2-5/8+1/16$ in the ten dimensions, one finds that  the following four-dilaton couplings must be added in the ten-dimensional theory:
\beqa
\frac{\gamma e^{-3\phi_0/2}}{\kappa^2}\cL_3( \phi)&=&\frac{\gamma e^{-3\phi_0/2}}{\kappa^2}(\frac{1}{16}\kappa^4)\bigg[\frac{3}{2}(\phi_{;hm}\phi^{;hm})^2-\phi_{;hn}\phi^{;hq}\phi_{;qr}\phi^{;rn}\bigg]
\eeqa
 which is the same as the four-dilaton couplings given by \reef{p41}. Note that up to Laplacian of dilaton and total derivative terms, one can write 
 \beqa
 (\phi_{;hm}\phi^{;hm})^2&=&-2\phi_{;m}\phi^{;hm}\phi_{;prh}\phi^{;pr}\nonumber\\
 &=&2\phi_{;mp}\phi^{;hm}\phi_{;rh}\phi^{;pr}+2\phi_{;m}\phi^{;hm}{}_{p}\phi_{;hr}\phi^{;pr}\nonumber\\
 &=&2\phi_{;mp}\phi^{;hm}\phi_{;rh}\phi^{;pr}+\frac{2}{3}(\phi_{;m}\phi^{;h}{}_{p} \phi_{;hr}\phi^{;pr})^{;m}\labell{iden}
 \eeqa
 The last term which is a total derivative can be dropped.
 
 The ten-dimensional theory which is reproduced by the string theory scattering amplitude  has then the following three parts:
 \beqa
 S= \frac{\gamma}{\kappa^2}\int d^{10}x\sqrt{-G} e^{-3\phi_0/2}[\cL_1(\bar{R})+\cL_2(\bar{R},\phi)+\cL_3(\phi)]\labell{S}
 \eeqa
 where the   Lagrangian $\cL_1(\bar{R})$ is the natural extension of the eight-dimensional Lagrangian  \reef{Y3} to ten dimensions, and the other two Lagrangians are the following:
 \beqa
\cL_2(\bar{R},\phi)&=&-\frac{\kappa^2}{4}\bigg[ 2 \bar{R}_{ m q  r  s } \bar{R}^{ pq  h s } \phi_{ ;h}{}^{m} \phi^{ ;r}{}_{p}+2 \bar{R}_{m  q  h  s } \bar{R}^{ p q r s } \phi^{ ;hm} \phi_{ ;rp}\nonumber\\
&& +2 \bar{R}_{  h {q}m  p } \bar{R}^{  q r m {p}} \phi^{ ;h}{}_{s} \phi_{ ;r}{}^s+\frac{1}{4} \bar{R}_{ h p m {r} } \bar{R}^{ h p m {r} } \phi_{ ;q}{}^{s} \phi^{ ;q}{}_s\bigg] \labell{RP} \\
\cL_3(\phi)&=& \frac{\kappa^4}{16}   (\phi_{; k}{}^n\phi^{;k}{}_n )^2  \nonumber
\eeqa 
Since the  action \reef{S} is reproduced by the string theory scattering amplitude, it is in the Einstein frame.

\section{On-shell T-duality}

In this section we would like to study the transformation of the action \reef{S} under the linear T-duality. Since the standard T-duality transformation rules are given in the string frame, we have to transform this action to the string frame. The relation between the string and the Einstein frames in ten dimensions is $G^s_{ab}=G_{ab}e^{\Phi/2}$. Using the perturbations \reef{perturb}, one finds this transformation at the linear level to be  the following:
\beqa
h_{ab}&=&h_{ab}+\eta_{ab}\phi/\sqrt{8}
\eeqa
where $h_{ab}$ on the left hand side is in the string frame. The right hand side is exactly the graviton and the dilaton in the replacement \reef{ltrans}.  Hence, the transformation of the linearized $\bar{R}_{ab cd}$  to the string frame is
\beqa
\bar{R}_{ab cd}=e^{-\phi_0/2}\cR_{ab cd}
\eeqa
where $\cR_{ab cd}$ is the following expression 
\beqa
\cR_{ab cd}&=&R_{ab cd}+H_{ab [c,d]}\labell{RH2}
\eeqa
One can check that the first term is symmetric under exchanging  the indices $ab\leftrightarrow cd$ whereas the second term is antisymmetric. 

The Einstein frame action \reef{S} then transforms to the  following action in the string frame:
\beqa
S &=&\frac{\gamma}{\kappa^2} \int d^{10}x\sqrt{-G}e^{-2\phi_0}[\cL_1(\cR)+\cL_2(\cR, \phi)+\cL_3(\phi)]\labell{Y5}
\eeqa
  The overall dilaton factor   $e^{-2 \phi_0}\sqrt{-G}$  is invariant under nonlinear T-duality. We are going to check the transformation of the other terms under the linear T-duality transformations \reef{TR} and \reef{TR2}. To this end, we have to do dimensional reduction on a circle and then study various components under these transformations. 
  
We begin by studying the T-duality in the absence of B-field and dilaton. Unlike the Lagrangian \reef{Y2},  the Lagrangian $\cL_1$ does not produce term with structure $RR_{yy}R_{yy}R_{yy}$ where $R_{yy}$ is the Riemann curvature in the string frame with two killing indices. So the action \reef{Y5} in the absence of the B-field and dilaton is invariant under off-shell T-duality transformations. Note that  the consistency with the T-duality \reef{TR} requires one to set to zero the Riemann curvature with one Killing index when the B-field is set to zero. 

Since   the Lagrangian $\cL_1(\cR)$ is invariant under the off-shell  T-duality in the absence of the B-field,  we hope this happens  in the presence of the B-field too.   The symmetries of the first term in \reef{RH2} are different from the symmetries of the second term, so we have to do the dimensional reduction for $\cL_1(R)$, $\cL_1(H)$, and mixed terms separately. We have found explicitly that the couplings with structure $H_yH_yR_yR_y$ where $H_y$ refers to $H_{ab[c,y]}$ and $R_y$ refers to   $R_{abcy}$,  are invariant under the linear T-duality \reef{TR}. We have also found that the couplings with structure $H_yH_yR_yR_y$ are invariant under the transformation \reef{TR}. Moreover, we have found the following transformations for other couplings:
\beqa
R_yR_yR_yR_y&\rightarrow &H_yH_yH_yH_y\nonumber\\
RRR_yR_y&\rightarrow&RRH_yH_y\nonumber\\
HHH_yH_y&\rightarrow&HHR_yR_y\nonumber\\
RR_yR_yR_{yy}&\rightarrow&-RH_yH_yR_{yy}
\eeqa
All above T-duality transformations have been verified without using any on-shell relations. So the Lagrangian $\cL_1(\cR)$ is invariant under the off-shell linear T-duality.

The Lagrangians  $\cL_2(\cR,\phi)$ and $\cL_3(\phi)$, however,  contain dilaton which is not invariant under the linear T-duality. The transformation \reef{TR2} indicates that the second derivative of the dilaton must be combined with the Ricci tensor to be invariant. The Ricci tensor, however, is zero using on-shell relations. So the second derivatives of the dilaton is invariant under on-shell linear T-duality. This immediately confirms that  $\cL_3(\phi)$ is invariant under the on-shell T-duality. Assuming that the second derivatives of the dilaton is invariant under T-duality, we have explicitly checked that all terms in the dimensional reduction of the Lagrangian  $\cL_2(\cR,\phi)$ transform among themselves under the transformation \reef{TR}. Therefore, the action \reef{Y5} is invariant under the on-shell linear T-duality.

\section{Off-shell T-duality}

In this section we are going to extend the action \reef{Y5} to be invariant under off-shell linear T-duality. The original gravity action \reef{Y1} if off by terms containing the Ricci and scalar curvatures. Including the B-field and dilaton through the replacement \reef{trans}, the theory does not include the divergence of the B-field strength and the Laplacian of the dilaton either. These missing terms however are all zero using on-shell relations. They can also be absorbed by the supergravity action \reef{super} in field redefinition $G\rightarrow G+\delta G$, $B\rightarrow B+\delta B$ and $\Phi\rightarrow \Phi+\delta\Phi $. This field redefinition, however, would change the standard form of the duality transformations   to receive higher derivative corrections. If one interested in the higher derivative action which is  invariant under the  standard form of the duality transformations, the action  must then take  a specific form for these couplings.   In this section we are going to include the Ricci curvature and the divergence of $H$ by requiring the action \reef{Y5} to be invariant under the standard linear T-duality transformations \reef{TR} and \reef{TR2}.

The Ricci curvature can easily be included in the action \reef{Y5} by noting that the combination of the second derivative of the dilaton and the Ricci curvature 
is invariant under the T-duality \reef{TR2}. So one has to use the replacement $2\sqrt{2}\kappa\phi_{;ab}\rightarrow R_{ab} +2\sqrt{2}\kappa\phi_{;ab}$ in the action \reef{Y5} to include the Ricci curvature. The resulting action produces many terms in the dimensional reduction which are invariant under T-duality. For example, the Lagrangian $\cL_2$ produces the  terms with structure of one Ricci curvature without Killing index, one Ricci curvature with two Killing indices and two $H$ each with one Killing index. They simplify to the following term:
\beqa
-4H_{ {mqy,s} }H_{ {hsy,q} }R_{ {hm} }R_{ yy }\labell{test}
\eeqa
It also produces the  terms with structure of one Ricci curvature without Killing index, one Ricci curvature with two Killing indices and two Riemann curvatures each with one Killing index. They also simplify to
\beqa
+4R_{ {hm} }R_{ yy}R_{ {mqsy} }R_{ {hsqy} }
\eeqa
which is the transformation of \reef{test} under the T-duality rules \reef{TR} and \reef{TR2}.

However, the resulting action is not yet fully invariant under T-duality because under the dimensional reduction one would find terms containing the Ricci curvature with one Killing index, $R_{ay}$. Under T-duality \reef{TR2} it goes to $-H_{ay}$. So we have to also include   the appropriate divergence of B-field strength to find the desired action. We propose the following replacement for the second derivative of the dilaton in the action \reef{Y5}:
 \beqa
2\sqrt{2}\kappa\phi_{;ab}&\rightarrow &\cH_{ab}\equiv R_{ab}+H_{ab}+2\sqrt{2}\kappa\phi_{;ab}\labell{Ricci}
\eeqa
Note that $H_{ab}$ is antisymmetric whereas the Ricci curvature and the dilaton term is symmetric.  

The couplings of two dilatons, one graviton and one B-field is zero, so the Lagrangian $\cL_2(\bar{R},\phi)$ is non-zero for two gravitons or two B-fields. As a result, we can write the third term in \reef{RP} as
\beqa
\bar{R}_{  h {q}m  p } \bar{R}^{  q r m {p}} \phi^{ ;h}{}_{s} \phi_{ ;r}{}^s&=&\bar{R}_{ m  p h {q}  } \bar{R}^{ m {p} q r } \phi^{ ;h}{}_{s} \phi_{ ;r}{}^s
\eeqa
 This makes no difference in the action \reef{S}. However, when one uses the  replacement \reef{Ricci}, the two sides of the above identity give different couplings.  Imposing the invariance under off-shell T-duality forces us to choose the left hand side of the above equation.  Similarly, the indices in  the second derivative of the dilaton are symmetric, so it makes no difference if one changes the  ordering of the dilaton indices in \reef{RP}. However, we have chosen the specific ordering which makes the extension of the action under the replacement \reef{Ricci} to be invariant under the off-shell T-duality.

Using the replacement \reef{Ricci} in the on-shell T-duality invariant action \reef{Y5} and taking the above points, we propose the following action:
 \beqa
S &=&\frac{\gamma}{\kappa^2} \int d^{10}x\sqrt{-G}e^{-2\phi_0}[\cL_1(\cR)+\cL_2(\cR, \cH)+\cL_3(\cH)]\labell{Y6}
\eeqa
where $\cL_1$ is given in \reef{Y3} and 
 \beqa
\cL_2(\cR,\cH)&=&-\frac{1}{2^5}\bigg[ 2 \cR_{ m q  r  s } \cR^{ pq  h s } \cH_{ h}{}^{m} \cH^{ r}{}_{p}+2 \cR_{m  q  h  s } \cR^{ p q r s } \cH^{ hm} \cH_{ rp}\nonumber\\
&& +2 \cR_{ m  p h {q} } \cR^{ m {p} q r } \cH^{ h}{}_{s} \cH_{ r}{}^s+\frac{1}{4} \cR_{ h p m {r} } \cR^{ h p m {r} } \cH_{ q}{}^{s} \cH^{ q}{}_s\bigg] \labell{RP3}\nonumber\\
\cL_3(\cH)&=&\frac{1}{2^{10}}  (\cH_{ k}{}^n\cH^{k}{}_n )^2 +\alpha\bigg[(\cH_{ k}{}^n\cH^{k}{}_n )^2-2\cH_k{}^n\cH^k{}_p\cH_s{}^n\cH^s{}_p\bigg] \nonumber
\eeqa
where $\alpha$ is a constant that our calculations can not fix it. It represents the identity \reef{iden} under the replacement \reef{Ricci}.  Obviously, the above action reduces to \reef{Y5} if one uses the on-shell relations $R_{ab}=H_{ab}=0$.  As we have shown before the Lagrangian $\cL_1(\cR)$ is invariant under off-shell linear T-duality. We have checked also that the Lagrangians $\cL_2(\cR,\cH)$ and $\cL_3( \cH)$ are invariant under the off-shell T-duality. For example, the Lagrangian $\cL_2(\cR,\cH)$ produces the following terms with structure of two Ricci curvatures each with one Killing index and two $H$ each with one Killings index:
\beqa
-2R_{hy}R_{ry}H_{mpy,h}H_{mpy,r}-4R_{hy}R_{py}H_{pqy,s}H_{hsy,q}+4R_{hy}R_{ry}H_{hsy,q}H_{rsy,q}\labell{test2}
\eeqa
It also produces the following terms with structure of two divergence of $H$  each with one Killing index and two Riemann curvatures each with one Killings index:
\beqa
-2H_{hy}H_{ry}R_{mphy}R_{mpry}-4H_{hy}H_{py}R_{pqsy}R_{hsqy}+4H_{hy}H_{ry}R_{hsqy}R_{rsqy}
\eeqa
It is the transformation of \reef{test2} under the T-duality rules \reef{TR} and \reef{TR2}. Similar examinations have been done for all   other structures.

\section{Discussion}

In this paper we have written the eight-dimensional Lagrangian \reef{Y3} found by Gross and Sloan \cite{Gross:1986mw} in a ten-dimensional form \reef{S}. We have shown that   the gravity and the B-field couplings in the  ten-dimensional  theory are invariant under off-shell linear T-duality, whereas, the dilaton couplings are invariant under on-shell  T-duality. We then impose the consistency of the dilaton couplings  with off-shell  T-duality to extend them to include   the couplings containing the Ricci curvature and the divergence of B-field strength \reef{Y6}. We have done this extension through the replacement \reef{Ricci} which is similar to the replacement \reef{trans} found in \cite{Gross:1986mw}. 

In the absence of B-field and dilaton, the actions \reef{Y6} and \reef{Y2} are identical up to some terms containing the Ricci and scalar curvatures \cite{Myers:1987qx} which can be eliminated by field redefinitions. These field redefinitions at the same time change the standard form of the T-duality rules \reef{nonlinear} to a non-standard form which contains higher derivative corrections. Since the action \reef{Y6} is invariant under the  standard linear T-duality, one would expect the supergravity with the Riemann curvature corrections  \reef{Y2} to be invariant under the non-standard T-duality transformation. In particular, the coupling \reef{non} may be canceled with the non-standard T-duality transformation of the supergravity \reef{super}. 

We have been able to find the couplings in \reef{Y1} which involve the Ricci curvature and the divergence of the B-field strength by imposing the off-shell T-duality of the effective action. However, the Lagrangian \reef{Y1}   has couplings involving the scalar curvature and the Laplacian of the dilaton.  How   can one find such couplings? We show that they can be found by imposing the linear S-duality and T-duality. In the type IIB theory, the couplings involving the divergence of the R-R three-form field strength can easily be included in  \reef{Y6}    by replacing $e^{-\phi_0}H_{ab }H_{cd}$ with the following S-duality invariant expression:
\beqa
e^{-\phi_0}H_{ab }H_{cd}\rightarrow e^{-\phi_0}H_{ab }H_{cd}+ e^{\phi_0}F_{ab }F_{cd}
  \nonumber
\eeqa
where $F_{ab}$ is  the divergence of the R-R three-form field strength. In the dimensional reduction of the resulting couplings, one finds the couplings which have $F_{ay}$. Under T-duality they produce the couplings in the type IIA theory which involve the divergence of the R-R two-form field strength $F_a$. Again in the dimensional reduction of the resulting theory, one finds the couplings which have $F_y$. Under T-duality theory produce the couplings in the type IIB theory which involve the Laplacian of the R-R scalar,  $F=C_{;a}{}^a$. Under the S-duality both the   dilaton and the R-R scalar appear in the  complex field $\tau=C+ie^{-\Phi}$, so one expects that there should be similar couplings for the Laplacian of the dilaton. On the other hand, the Laplacian of the dilaton is not invariant under linear T-duality. The combination of the  scalar curvature and the Laplacian of the dilaton, $R+4\Phi_{;a}{}^a$, is invariant under the T-duality \cite{Garousi:2012yr}. So in the couplings involving the Laplacian of the dilaton one should replace $4\Phi_{;a}{}^a\rightarrow R+4\Phi_{;a}{}^a$ to find the couplings involving the scalar curvature.

The linear   T-duality invariant action \reef{Y6} includes all couplings of four  B-fields at order $\alpha'^3$.  At this order, however, there are higher order couplings, \eg couplings involving $H^8$. These couplings may be found by extending the linear T-duality of \reef{Y6} to nonlinear T-duality. The appropriate framework for such studies is the double field theory formalism \cite{Hull:2009mi,Hohm:2010jy}. In this formalism a T-duality invariant action is expressible  in terms of the generalized metric \cite{Hohm:2010pp}
\beqa
\pmatrix{G_{ij}-B_{ik}G^{kl}B_{lj}& B_{ik}G^{kj} \cr 
  -G^{k}B_{kj}& G^{ij} }\labell{metric}
  \eeqa 
which transforms covariantly under the T-duality, and the T-duality invariant field $d$ which is related  to the dilaton via the field redefinition $e^{-2d}=\sqrt{-G}e^{-2\Phi}$. The double field theory action must be invariant under the large gauge transformation which is a combination of the conventional diffeomorphism and the B-field gauge transformations \cite{Hull:2009mi,Hohm:2010jy}. In one sets to zero the nonlinear terms in the above metric, then one may be able to rewrite the action \reef{Y6} in terms of the linearized metric. That action might then  automatically include the higher order couplings of B-field by restoring  the nonlinear terms in the metric. 

Finally, let us compare the $\alpha'^3$-order spacetime action \reef{Y6} with the string frame $\alpha'^2$-order D-brane action    found in \cite{Bachas:1999um,Garousi:2009dj,Garousi:2011fc}
\beqa
S&\sim& \int d^{p+1}x\,\sqrt{-G}e^{-\phi_0}\bigg[-2(\hat{R}_{ab}+ \Phi_{;ab})(\hat{R}^{ab}+ \Phi^{;ab})+2(\hat{R}_{ij}+ \Phi_{;ij})(\hat{R}^{ij}+ \Phi^{;ij})\nonumber\\
&&+R_{abcd}R^{abcd}-R_{abij}R^{abij}-\frac{2}{3}\prt_aH_{ijk}\prt^aH^{ijk}-\frac{4}{3}\prt_iH_{abc}\prt^iH^{abc}+2\prt_aH_{bci}\prt^aH^{bci}
\bigg]\nonumber
\eeqa
In this equation the indices $a,b,c,d$ are the world volume indices and $i,j,k$ are the transverse indices. Note that the Ricci curvature $\hat{R}$ is the Riemann curvature that is contracted only with the world volume metric \cite{Bachas:1999um}. This makes the combination $\hat{R}_{\mu\nu}+ \sqrt{2}\kappa\phi_{;\mu\nu}$ to be invariant under linear T-duality.   The above action is invariant under off-shell linear T-duality \cite{Garousi:2009dj,Garousi:2011fc}. In both the D-brane  action and the spacetime action \reef{Y6}, the dilaton appears via extending the corresponding Ricci curvature. One may expect the same thing would happen for higher $\alpha'$-order curvature couplings. While the B-field couplings appear in the spacetime action \reef{Y6} via extending the Riemann curvature \reef{RH2} and the Ricci curvature \reef{Ricci}, in the D-brane theory there is no such an analogy. 
 


\end{document}